\documentclass[11pt,preprintnumbers,nofootinbib,prd]{revtex4}

\usepackage[hypertex]{hyperref}
\usepackage{amsmath,amssymb}	
\usepackage{graphicx}
\usepackage{slashed}
\def\lesssim{\mathrel{\mathpalette\vereq<}}

\begin{document}

\pagestyle{plain}

\title{Lepton Number Violating Signals of the Top Partners in the Left-Right Twin Higgs Model}

\author{Hock-Seng Goh}
\affiliation{Department of Physics, University of California,
Berkeley, CA}

\author{Christopher A. Krenke}
\affiliation{Department of Physics, University of Maryland, College Park,
MD 20742}
\affiliation{Department of Physics, University of Arizona, Tucson, AZ 85721}




\begin{abstract}


We study the collider signatures of the left-right twin Higgs model in the case
that the right-handed neutrino mass is less than the mass of the right-handed
gauge boson.  In this scenario, new leptonic decay chains open up, allowing the
particles which cancel the one-loop quadratic divergences of the Higgs, the
right-handed gauge bosons and top-partners, to be discovered.  Half of these
events contain same-sign leptons without missing energy, which have no genuine
standard model background and for which the backgrounds are purely
instrumental.   These signals may be used to complement other collider
searches, and in certain regions of parameter space,  may be the only way to
observe the particles responsible for natural electroweak symmetry breaking in
the left-right twin Higgs model.


\end{abstract}

\pacs{} \maketitle

\newpage 

\section{Introduction}

The standard model (SM) with a fundamental Higgs field suffers from an extreme
sensitivity to short distance physics.  If the cutoff of the SM is taken to be
the Planck scale, this sensitivity leads to a tremendous fine tuning of the
dimensionful parameters in the Higgs potential and a large hierarchy between
the weak and Planck scales.  If the cutoff of the SM is taken to be about
5~TeV,  the minimum allowed by precision electroweak data, an unnatural
adjustment of parameters persists and results in a ``little hierarchy''
\cite{Barbieri}. This fact implies that new physics should exist at the TeV
scale which is responsible for resolving the little hierarchy problem. 
This is an important observation, as the Large Hadron Collider (LHC) at CERN
has been built with the aim of detecting particles with TeV scale masses.
Therefore, new physics that is tightly connected to the nature of electroweak
symmetry breaking is expected to be within the reach of the LHC. The nature of
the new physics that cures the hierarchy problem, or the little hierarchy
problem if the ultraviolet cutoff is taken to be less than 5~TeV, is highly
constrained. Electroweak precision measurements have imposed very strong bounds
on any new physics around a TeV. These constraints pose a great challenge to
designing models meant to address the little hierarchy problem.

One class of theories that address the little hierarchy problem is known as the
little Higgs \cite{Little1,Little2,Tparity}. In these models, the SM Higgs
doublet is a pseudo-Nambu-Goldstone boson (PNGB) of some spontaneously broken
approximate global symmetry in which the SM $SU(2)$ electroweak symmetry is
embeded \cite{GP,KG}. The Higgs mass vanishes at tree level due to shift
symmetry, but will be generated by radiative corrections when interactions that
break the global symmetry, such as gauge and Yukawa interactions, are included.
At one-loop, multiple approximate global symmetries restrict the form of the
quadratic terms in the Higgs potential such that together they form an invariant
of the full global symmetry.  
Therefore, the one loop Higgs mass is only logarithmically sensitive to UV
physics.  This class of models is able to stabilize the elctroweak scale
against the UV cutoff up to a scale of about 5 - 10 TeV.

Another class of theories that solves the little hierarchy problem by
identifying the Higgs as a PNGB are twin Higgs models
\cite{twin,lrtwin,littletwin}.  Instead of protecting the Higgs mass from
receiving large radiative corrections by using several approximate global
symmetries, twin Higgs theories use a discrete symmetry in combination with an
approximate global symmetry to eliminate the quadratic divergences that arise
at loop level.  Together with the gauge symmetries of the model, the discrete
symmetry mimics the effect of a global symmetry, thus stabilizing the Higgs
mass.

In the left-right Twin Higgs model \cite{lrtwin}, the SM gauge symmetry is
extended to $SU(2)_L \times SU(2)_R \times U(1)_{B-L}$ \cite{originalLR}, which
is embedded into a global $U(4)$ symmetry.  The Higgs arises as a PNGB when the
$U(4)$ symmetry is spontaneously broken to $U(3)$.  An additional $Z_2$ ``twin
symmetry'' ensures that the quadratic terms in the Higgs potential have an
accidental $U(4)$ symmetry.  Since $U(4)$ invariant terms cannot contribute to
the potential for the Goldstones, the Higgs is protected from receiving
quadratically divergent contributions to its mass parameter.  To evade
precision electroweak bounds on $SU(2)_R$ gauge bosons without significantly
affecting naturalness, an additional Higgs field, $\hat{H}$, is introduced that
transforms as a fundamental under a second global $U(4)$ symmetry.  This
addition makes the approximate global symmetry of the theory $U(4) \times
U(4)$.  The new global symmetry does not significantly alter the form of the SM
Higgs potential, allowing electroweak symmetry breaking to still happen
naturally. 

%

To identify the twin mechanism it is important to observe the heavy top quark
partner, $T_H$, and the right-handed gauge boson, $W_R$.  For a reasonable
choice of parameters, the most straightforward way to observe both of these
particles involves decays of the heavy top quark, which has a channel
containing final state leptons that can be used as a trigger.  It may be
possible to reconstruct these events and observe the heavy top quark at the LHC
\cite{shufang}.  However, this decay channel depends on a free parameter $M$,
which could be very small or zero.  In this limit, the heavy top quark can only
decay hadronically \cite{shufang}, making these particles very difficult to
observe at the LHC due to the large QCD background. 

In this paper, we study an alternative way to observe the heavy top quark and
the right-handed gauge boson.  If a TeV scale right-handed Majorana neutrino is
realized in the left-right twin Higgs model such that $m_{\nu_R} < m_{W_R}$,
new leptonic channels open up that may allow detection of $W_R$ and $T_H$ at
the LHC.  Moreover, because the right-handed neutrino is Majorana, half of
these decays are lepton number violating same-sign dilepton events without
missing energy, which has no genuine SM background.  If $M$ is small or zero,
these lepton number violating signals  may be the only way to observe the heavy
top quark and the right-handed gauge boson at the LHC.

This paper is organized as follows:  In section II, we review the left-right
twin Higgs model and discuss its phenomenology in the decoupling case where the
parameter $M$ is set to zero. In section III, we implement neutrino masses into
the model and discuss 
a TeV scale right-handed neutrino.  We study  the collider phenomenology of the
model in section IV, focusing on searches for the right-handed gauge boson,
$W_R$, and the heavy top partner, $T_H$. We then conclude in section V.

\section{Left-right Twin Higgs model}

\subsection{Matter Content}

The fermionic content of the left-right twin Higgs model contains three
generations of
\begin{align}
Q_L & = \left(u,d \right)_L =\mathbf{(2,1,1/3)} \; \;
L_L = \left(\nu,e \right)_L = \mathbf{(2,1, -1)} \nonumber \\
Q_R &= \left(u,d \right)_R = \mathbf{(1,2, 1/3)} \; \; L_R =
\left(\nu,e \right)_R = \mathbf{(1,2, -1)}, 
\end{align}
where the square brackets indicate the quantum numbers of the corresponding
fields under the $SU(2)_L \times SU(2)_R \times U(1)_{B - L}$ gauge symmetry of
the theory. We see that in addition to the SM fermions, the theory includes
right-handed neutrinos as required by left-right symmetry.
There are two sets of Higgs fields which have quantum numbers
\cite{Atlr,RabiLR}
\begin{align}
H_L &= \mathbf{(2, 1, 1)} \; \; \; \; \; \; H_R = \mathbf{(1,2,1)} \nonumber\\
\hat{H}_L &= \mathbf{(2, 1, 1)} \; \; \; \; \; \; \hat{H}_R =
\mathbf{(1,2,1)}.
\end{align}
The reason for introducing the extra set of Higgs fields $\hat{H}$ is to
satisfy precision electroweak constraints on $SU(2)_R$ gauge bosons.  These
constraints require the symmetry breaking scale $f$ of $SU(2)_R$ to be larger
than about 2 TeV \cite{Kingman}.  However, for this value of $f$, contributions
to the Higgs potential from the top sector are very large since the top Yukawa
is order one.  This effect tends to reintroduce fine tuning to the model,
destabilizing the weak scale.  
By adding an additional Higgs field $\hat{H}$, which acquires a vev  $\langle
\hat{H} \rangle = \hat{f} \sim 2$~TeV and does not couple to fermions,
precision electroweak constraints on $SU(2)_R$ gauge bosons can be satisfied
without affecting the top sector.  This arrangement can be justified by
imposing a discrete symmetry under which $\hat{H}$ is odd while all other
fields are even.  This symmetry allows $\hat{H}_L$ to be stable, making it a
natural dark matter candidate.  It has been shown that this can account for the
observed relic abundance of dark matter \cite{darkmatter}.

%
%

The Higgs potential is assumed to have an approximate $U(4) \times U(4)$
symmetry of which the $SU(2)_L \times SU(2)_R \times U(1)_{B - L}$ sub-group is
gauged.
After breaking the global $U(4)$ and the gauged $SU(2)_R$ symmetries, the SM
Higgs doublet, which is among the NGBs, has no potential at tree level.
However, a potential for the Higgs will be radiatively generated at one loop.
In this scenario, both the mass and the quartic coupling of the Higgs are loop
suppressed. To further reduce fine tuning, a tree level Higgs quartic can be
introduced without generating a corresponding tree level mass term for the
Higgs, as discussed in \cite{littletwin}. 
Since the Higgs potential is not relevant to our discussion of neutrino masses
or collider signals, we shall not go into a detailed discussion of the Higgs
potential.

The down-type Yukawa couplings of the SM emerge from non-renormalizable
couplings of the form
\begin{equation}
\left(\frac{\overline{Q}_R H_R H_L^{\dagger} Q_L \; \; + \;\;
\overline{L}_R H_R H_L^{\dagger} L_L}{\Lambda} \right)  \; \; + \;
\; {\rm h.c.,}
\end{equation}
while the up-type Yukawa couplings of the SM emerge from non-renormalizable
couplings of the form
\begin{equation}
  \left( \frac{\overline{Q_R} H_R^{\dagger} H_L Q_L}{\Lambda} \right) \; \; + \;\;
{\rm h.c.}
\end{equation}
When the field $H_R$ acquires a VEV of order $f$ breaking $SU(2)_R \times
U(1)_{B - L}$ down to $U(1)_Y$, these non-renormalizable couplings reduce to
the familiar Yukawa couplings of the SM.  Unfortunately, this method of
generating SM Yukawa couplings does not work well in the top sector since the
top Yukawa coupling is order one.  This problem is remedied by introducing the
following vector like quarks, which transform as 
\begin{equation}
  T_L = \mathbf{(1,1,4/3)} \ \ \  T_R = \mathbf{(1,1,4/3)}
  \label{vectorquarks}
\end{equation}
under $SU(2)_L \times SU(2)_R \times U(1)_{B-L}$.  We can then write the
following left-right symmetric interactions
\begin{equation}
    \left( y \overline{Q}_R H^{\dagger}_R T_L + 
    y \overline{Q}_L H^{\dagger}_L T_R + M \overline{T}_L T_R \right) + {\rm h.c.}   
  \label{LRvectorquarklagran}
\end{equation}
The right-handed top quark of the SM then emerges as a linear combination of
$T_R$ and the third generation up-type quark in $Q_R$, while the orthogonal
linear combination is heavy. Provided $M \lesssim f$ and $y$ is of order one
the physical top Yukawa will then also be of order one. 

The parameter $M$ controls the mixing of the left-handed top with the SU(2$)_L$
singlet $T_L$, and is therefore constrained by $Z \rightarrow b \; \bar{b}$.
However, nothing prevents $M$ from simply being set to zero and therefore this
is not a particularly tight constraint. However, the collider phenomenology of
this model will depend on the size of this parameter. As we will see below,
when $M$ is small, with $M=0$ as a extreme case of this scenario, the heavy top
becomes difficult to observe in a hadron collider since it decays dominantly
into an all jet final state.

\subsection{Phenomenology}
\label{dark}

The left-right twin Higgs contains many new particles which may be observable
at the LHC.  The new particles include the right-handed gauge bosons $W_R$ and
$Z_R$, a heavy top quark $T_H$, a right-handed neutrino $\nu_R$,  and the
Higgses $\hat{h}^T = (\hat{h}^+, \hat{h}^0)$, $\phi^{\pm}$ and $\phi^0$.  The
gauge boson masses depend on the larger vev $\hat{f} $ and range from about 1 -
4~TeV, while the heavy top is typically lighter, ranging from 0.5 - 1~TeV.  The
$\phi^0$ mass depends on a free parameter in the theory, but is usually taken
to be about 100~GeV.  The charged Higgs $\phi^{\pm}$ mass ranges from about 200
- 400~GeV, while the $\hat{h}$ mass ranges from about 300 GeV to 1 TeV.  The
right-handed neutrino mass arises from the operator 
\begin{align}
\left( \frac{{L}_R  {\hat{H}}_R {\hat{H}}_R
L_R~+~{L}_L  {\hat{H}}_L {\hat{H}}_L L_L}{\Lambda} \right)
\label{nu_op}
\end{align}
and is of order $\hat{f}^2/\Lambda$, which is about 1.5 TeV for
$\hat{f} \sim 4$~TeV and $\Lambda \sim 10$~TeV.  We will have more to say about
neutrino masses in section \ref{TeVseesaw}.

What are the collider signatures of this model?  The $Z_R$ decays to leptons
providing a very clean signal, which may be observable at the LHC
\cite{shufang}.  Detection of $W_R$ however, is more subtle.   For now assume
that $m_{\nu_R} > m_{W_R}$ and leptonic decays of the $W_R$ are kinematically
forbidden.  This was the scenario studied in \cite{shufang}.  In this case, the
$W_R$ decays a large fraction of the time (20\% - 30\%) to a heavy top and a
$b$-jet.  Therefore the discovery potential of both the heavy top and $W_R$
depend critically on how the heavy top decays. 
As discussed in \cite{shufang}, the heavy top is produced in association with a
$b$-quark, with a production cross section of about 500~fb.  For a reasonable
choice of $M = 150$~GeV, $T_H$ decays most often to $\phi^{\pm} b$.  For
this value of $M$, the $\phi^{\pm}$ then decays mostly to $tb$. It is then
possible to trigger on the  leptonic decay of the top, giving the following
decay chain
\begin{align}
  T_H \rightarrow \phi^{\pm} b \rightarrow t b b \rightarrow W b b b \rightarrow l \nu b b b. 
  \label{THdecaychain}
\end{align}
This scenario has been studied and shown to be detectable at the LHC with total
luminosity of 10~fb$^{-1}$ \cite{shufang}.

%

\subsubsection*{$M=0$: The Dark Side of the Model}


\begin{figure}
 \begin{minipage}{0.47\linewidth}
\begin{center}
\resizebox{2.5in}{!}{\includegraphics*[265,600][435,725]{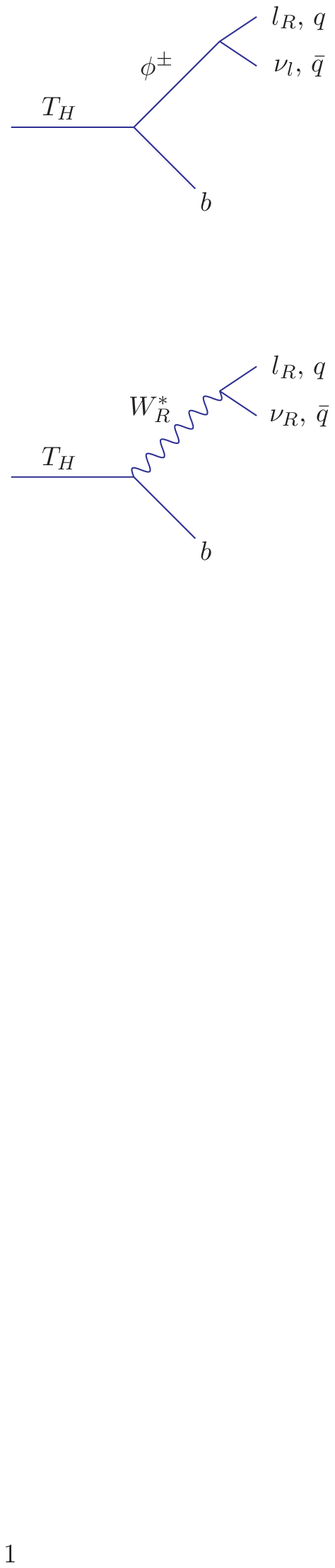}}
\end{center}
 \end{minipage}
  \begin{minipage}{0.515\linewidth}
\begin{center}
\resizebox{2.5in}{!}{\includegraphics*[275,460][440,580]{Tdecay.ps}}
\end{center}
 \end{minipage}
\caption[Possible $T_H$ decay ($M$=0)]{Possible decays of the heavy top in the limit that $M = 0$.} 
\label{fig:T-decay}
\end{figure}

In the previous section, we reviewed the phenomenology of the left-right twin
Higgs model when $m_{\nu_R} > m_{W_R}$ and for a small but reasonable value of
$M = 150$~GeV.  In this case,  the decays of the heavy top included a leptonic
final state, which could be used as a trigger.   However, the phenomenology
changes significantly when $M$ is very small, less than about 10~GeV.  The
crucial difference in this case is the decay of $\phi^{\pm}$, which previously
decayed to a SM top quark which then decayed leptonically.  When $M=0$,  the
$\phi^{\pm}$ decays purely to charm and strange, leading to an all jet final
state for heavy top decay. 

The reason $\phi^{\pm}$ does not decay to the SM top quark can be understood as
follows.  If $\phi^{\pm}$ is thought of as the charged component of $H_R$, then
$\phi^{\pm}$ couples directly only to $b_R$ and $T_L$, which in the limit that
$M=0$ are identified as the right-handed SM $b$ quark and the left-handed heavy
top, respectively.  When $M \neq 0$, mixing between $T_L$ and $T_R$ induces a
coupling to the SM top quark that is proportional to $M/f$ for $M << f$.
Therefore, in the limit $M \rightarrow 0$, $\phi^{\pm}$ cannot decay to a SM
top quark.

What about other decay channels for $T_H$? 
Two other decay channels are possible and are shown in Fig.~\ref{fig:T-decay}.
In the first decay channel, $\phi^{\pm}$ will dominantly decay to $q\bar{q}$
because the leptonic decay channel is suppressed by the neutrino Yukawa
coupling constant. In the second decay channel, $\nu_R$ is kinematically
unavailable and therefore the leptonic channel is only available through an off
shell $\nu_R$, which is highly suppressed. Therefore, in the limit that $M
\rightarrow 0$,  $T_H$ can only decay hadronically, leading to an all-jet final
state for heavy top decay.  
Detection of the heavy top at the LHC then becomes difficult. In this scenario,
the model may become one of those in which the true mechanism of natural
electroweak symmetry breaking is beyond the reach of the LHC.

Since the small $M$ parameter space is large, technically natural, and does not
affect the twin mechanism, it is important to examine this possibility more
closely. The hope lies in the size of right-handed neutrino mass relative to
the mass of the $W_R$.  If $m_{\nu_R} < m_{W_R}$, leptonic decay channels of
the heavy top open up and provide a way to observe the heavy top partner that
is independent of the parameter $M$.  As a preliminary, we discuss 
neutrino mass generation in the left-right twin Higgs model
in the next section.

\section{Neutrino Mass: Seesaw at the TeV Scale}
\label{TeVseesaw}
There is more than one way to implement neutrino mass in the left-right twin
Higgs model.  For a detailed study of neutrino mass generation in this context,
see \cite{Hidalgo}.  If lepton number is a good symmetry of the theory, the
neutrino masses must be Dirac.  In this case, the smallness of the neutrino
masses can be understood purely as a result of their small Yukawa couplings.
If lepton number is not conserved, the neutrino masses can be Majorana.  In
this case, the lightness of the SM neutrinos can be understood as a result of
the seesaw mechanism \cite{seesaw}.    
We will assume that left-right symmetry is exact in the neutrino sector as in
all other sectors of the model.  The most general collection of operators that
generate neutrino masses are the following: Dirac neutrino masses arise from
the operators
\begin{eqnarray}
  y_{\nu}\left(\frac{L_R H_R H_L L_L}{\Lambda} \right) \; + \; {\rm
h.c.} & \rightarrow & y_{\nu}\frac{fv}{\Lambda}\nu_L \nu_R \; + \; {\rm
h.c.}\\\nonumber &=& m_D \;\nu_L \nu_R \; + \; {\rm h.c.},
\end{eqnarray}
while the operators
\begin{align}\label{eq:majorana}
y_1 \left(\frac{{L}_R \hat{H}_R \hat{H}_R L_R~+~{L}_L  \hat{H}_L \hat{H}_L L_L}{\Lambda} \right) 
\; + \; {\rm h.c.} &\rightarrow y_1\frac{\hat{f}^2}{\Lambda}\nu_R\nu_R\; + \; {\rm h.c.} \\ 
y_2 \left( \frac{{L}_R  {H}_R {H}_R L_R~+~{L}_L  {H}_L {H}_L L_L}{\Lambda} \right)\; +
\; {\rm h.c.} & \rightarrow y_2 \left( \frac{{f}^2}{\Lambda}\nu_R\nu_R + 
\frac{{v}^2}{\Lambda}\nu_L\nu_L \right) \; + \; {\rm h.c.}
\label{eq:majorana2}
\end{align}
generate Majorana masses for the right-handed neutrinos, $\nu_R$, and the left
handed neutrinos, $\nu_L$. One possibility is that we assume lepton number is
not violated. In this case, the operators in eq.~(\ref{eq:majorana}) and
eq.~(\ref{eq:majorana2}) are not present. The light neutrinos,
$\nu=(\nu_L,\bar{\nu}_R)$, are Dirac fermions and the small neutrino masses are
just the result of small Yukawa couplings, which are around $10^{-12}$. The
other possibility is that light Majorana neutrinos are generated through a TeV
scale seesaw mechanism.  If we no longer assume lepton number conservation, all
the operators above are allowed. Since the operator in eq.~(\ref{eq:majorana2})
gives both $\nu_R$ and $\nu_L$ a Majorana mass, this term should be small, i.e.
$y_2 < 10^{-11}$.  Note that eq.~(\ref{eq:majorana}) does not generate a Majorana mass for
$\nu_L$ because $ \langle \hat{H}_L \rangle = 0$.  The SM neutrinos
can then obtain a Majorana mass of the right size if the coupling constant $y_{\nu}$
is $\sim 10^{-5}$, which is of order the electron Yukawa coupling.  We will
follow this possibility from now on and assume $y_2 = 0$.  A $Z_4$ symmetry
where $H$ is neutral may be used to justify this possibility.

\subsection*{Constraints on Majorana Right-handed Neutrinos}

Right-handed neutrinos with masses of order a TeV have been studied by
several authors \cite{RHneutrino}.  Since there are three generations of
right-handed neutrinos, the details of experimental constraints and their
collider signals depend substantially on their mass differences and mixing
angles. For simplicity, we do not consider the most general mass matrix and
assume a nearly degenerate mass spectrum. 

What are the constraints on right-handed neutrinos?  Big Bang nucleosynthesis
(BBN) puts severe constraints on new light degrees of freedom.  However,
particles that are heavier than an MeV and which do not decay during the era of
BBN are completely free of this constraint.  
Another bound comes from tritium decay, but that also only constrains light
particles with masses less than about an MeV.  There are stronger restrictions
on massive right-handed neutrinos from precision measurements of $Z$-decay and
single $\nu_R$ production \cite{L3small,L3large}.  However, we will only
consider right-handed neutrino masses of order a few hundred GeV, which are
also free from these constraints.


If the right-handed neutrino is Majorana, the most stringent bound on its mass
arises from neutrinoless double beta decay. The bound can be approximately
expressed in terms of the bound on the light neutrino effective mass
$m_{ee}^{max}$ as in \cite{0vbbth}
\begin{equation}
    \frac{m_{\nu_R}}{p^2-m_{\nu_R}^2}\prod_{i=1,2} \frac{V_{i,q}V_{i,e}}{g_2^2}
    \left( \frac{m_W^2}{m_{xi}^2} \right) \leq \frac{m_{ee}^{max}}{m_em_p}\frac{1-\chi_F}{\chi_H},
\label{betadecay}
\end{equation}
where $m_{xi}$ are masses of the particles that mediate beta decay and
$V_{i,q}$ and $V_{i,e}$ are the corresponding couplings to the quarks and the
electron, respectively. $p$ is the typical energy exchanged in the process,
which is of order 100 MeV. $\chi_{F}$ and $\chi_{H}$ are the nuclear matrix elements
corresponding to exchanging $\nu_{L}$ and $\nu_R$, respectively, and are given in
\cite{0vbbth}.
For example, in an extension of the SM with only right-handed neutrinos and the
seesaw mechanism, $x_i=W^{\pm}$ and the couplings are $V_{i,e}\sim g_2
\delta_{ss}$ where $\delta_{ss}$ is the seesaw mixing factor, $\delta_{ss} \sim
\frac{m_D}{m_{\nu_R}} \sim \sqrt{\frac{m_{\nu}}{m_{\nu_R}}}$.  For a TeV scale
right-handed neutrino, $\delta_{ss} \sim 10^{-7}$.  In general, several
diagrams may contribute to neutrinoless double beta decay and various
parameters will be constrained by experiment. In the case of the left-right
twin Higgs model, the diagrams with standard model $W$ exchange contain the
seesaw mixing factor and, therefore, are much more suppressed than those with
$W_R$ exchange, which are only suppressed by the mass of the $W_R$.  Other
subdominant diagrams with charged Higgs exchange are also suppressed. The
contribution from $W_R$ to neutrinoless double beta decay therefore leads to
the tightest constraint.  Using the experimental bound given in \cite{0vbbex}
and eq.~(\ref{betadecay}), we find
\begin{equation}
    m_{\nu_R} {m^4_{W_R}} \geq 1.56 \ \text{TeV}^{5}.
\end{equation}
The lower bound of the right-handed neutrino mass is then
\begin{equation}
  {m_{\nu_R}} \geq 120 \left( \frac{1.9 \  \text{TeV}}{m_{W_R}} \right)^4 \ \text{GeV},
\end{equation}
which is well below the range of mass we will be considering.  We will from now
on treat the neutrino mass, $m_{\nu_R}$, as a free parameter and study its
collider phenomenology.

\section{$M = 0$ Phenomenology} The phenomenology of the left-right twin Higgs
model has been studied by many authors \cite{shufang,phenoLRTH}.
We will focus on the limit where the top mixing parameter, $M$, is set to zero.
When $M =0$ and the right-handed neutrino is heavier than $W_R$, some of the
new particles including $W_R$ and heavy top partner $T_H$ are difficult to
detect because their decay channels are dominated by hadronic final states.
Here we consider the limit when the right-handed neutrino mass is less than the
mass of $W_R$. The search for $W_R$ will then be much more effective due to the
opening of leptonic decay channels. Even better, the leptonic decay has a 50\%
chance of violating lepton number due to the Majorana nature of $\nu_R$. The
same advantages will also apply to $T_H$, but these searches depend on the mass
of $\nu_R$ relative to the mass of $T_H$.  There are  two possibilities: (i)
$m_{\nu_R}<m_{T_H}$ and (ii) $m_{\nu_R}>m_{T_H}$, which we consider separately.
In the following analysis we choose the following typical parameter set: $f =
800$~GeV, which implies $\hat{f} \approx 4$~TeV, $m_{W_R} \approx 1.9$~TeV and
$T_H \approx 780$~GeV for a reasonable choice of soft parameters.  A different
choice of soft parameters will lead to a different set of masses, so these are
not strict mass relations.  However, this will not qualitatively affect our
conclusions.  Let us begin with a discussion of the search for $W_R$, which can
be done independently from $T_H$.

\subsection{$W_R$ Search} 
\begin{figure}
\begin{center}
\resizebox{2.5in}{!}{\includegraphics*[250,610][430,740]{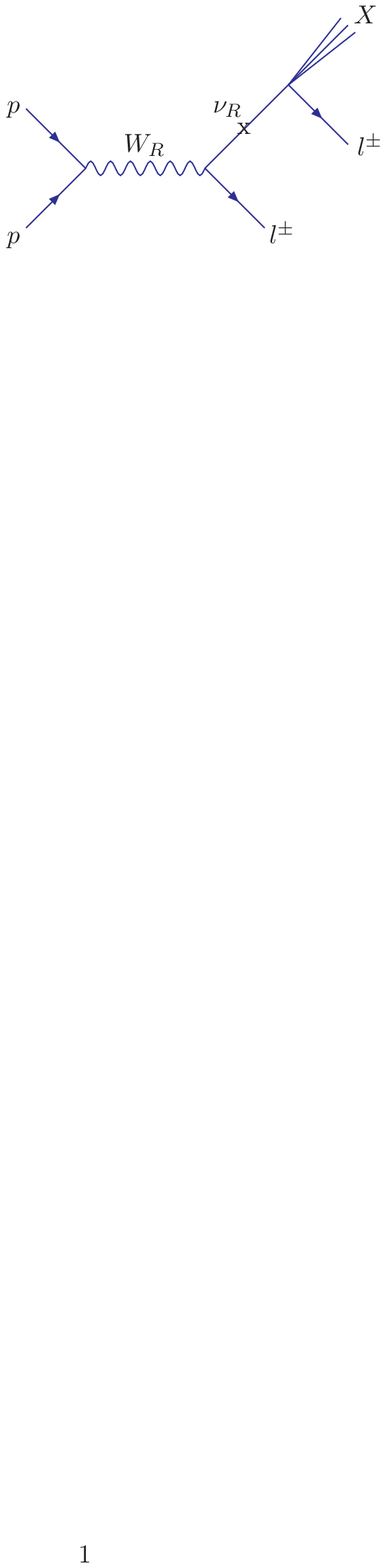}}
\end{center}
 \caption[$W_R$ production and decay via $\nu_R$.]{Diagrammatic view of $W_R$ production and its lepton number violating decay channel.} \label{fig:Wsearch}
\label{WRsearch}
\end{figure}

The $W_R$ is dominantly produced via a Drell-Yan
process and subsequently decays leptonically to $\nu_R + l^{\pm}$ with a
branching fraction of about a 10\%.  $\nu_R$ then decays to $l^{\pm} + X$
through an off shell $W_R$ or an on shell charged Higgs $\phi^{\pm}$, as shown
in Fig.~\ref{WRsearch}.  Here, $X$ represents any number of final state jets.
Due to the fact that $\phi^{\pm}$ only decays hadronically and the leptonic
decays of the off shell $W_R$ are kinematically forbidden, $X$ cannot contain
any leptons. 
As argued in the section \ref{dark}, this is precisely why the heavy top decays
purely hadronically in the decoupling limit when $M \rightarrow 0$.
Since $\nu_R$ is Majorana, half of these events will contain same-sign leptons.
%
%
%
The signal is therefore same-sign dilepton $l^{\pm}l^{\pm}X$ events without
missing energy, which has no genuine SM background.

\begin{figure}[t]
\centering
\includegraphics*[width =  0.45\textwidth]{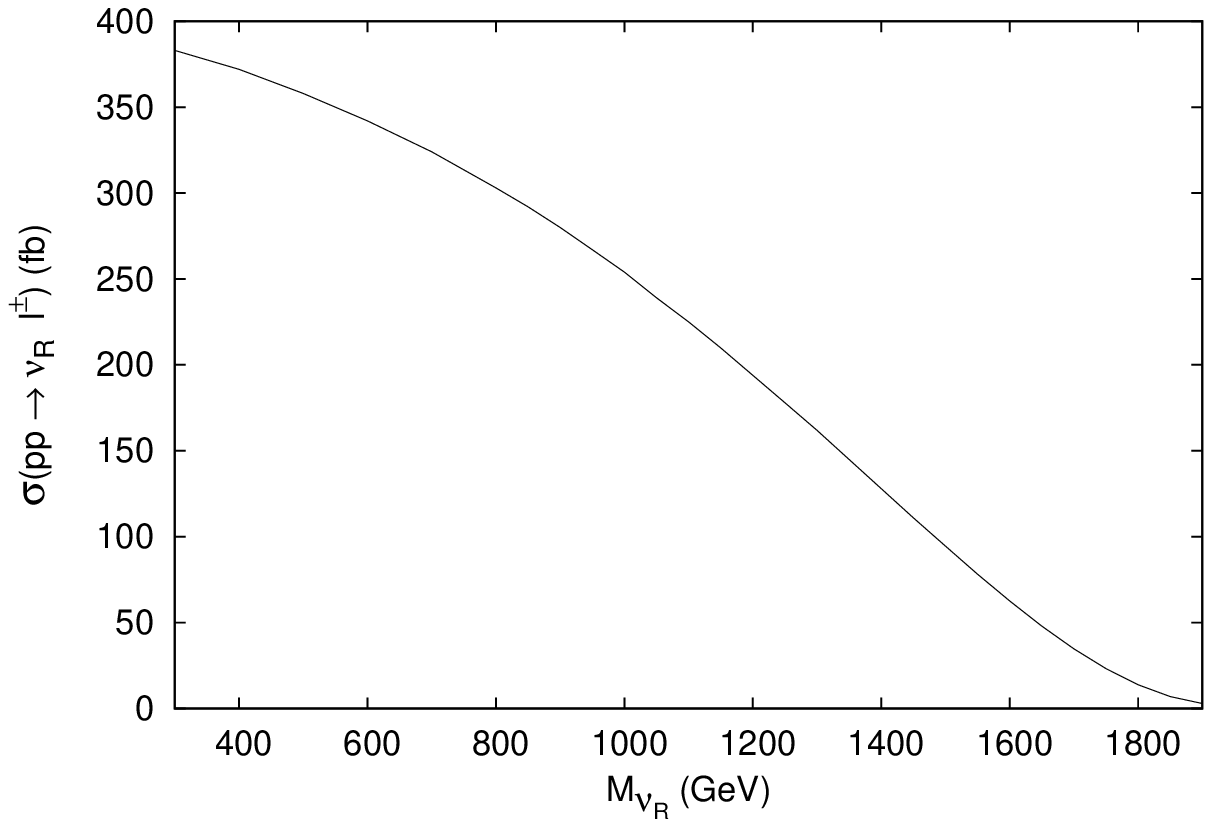}
\includegraphics*[width =  0.45\textwidth]{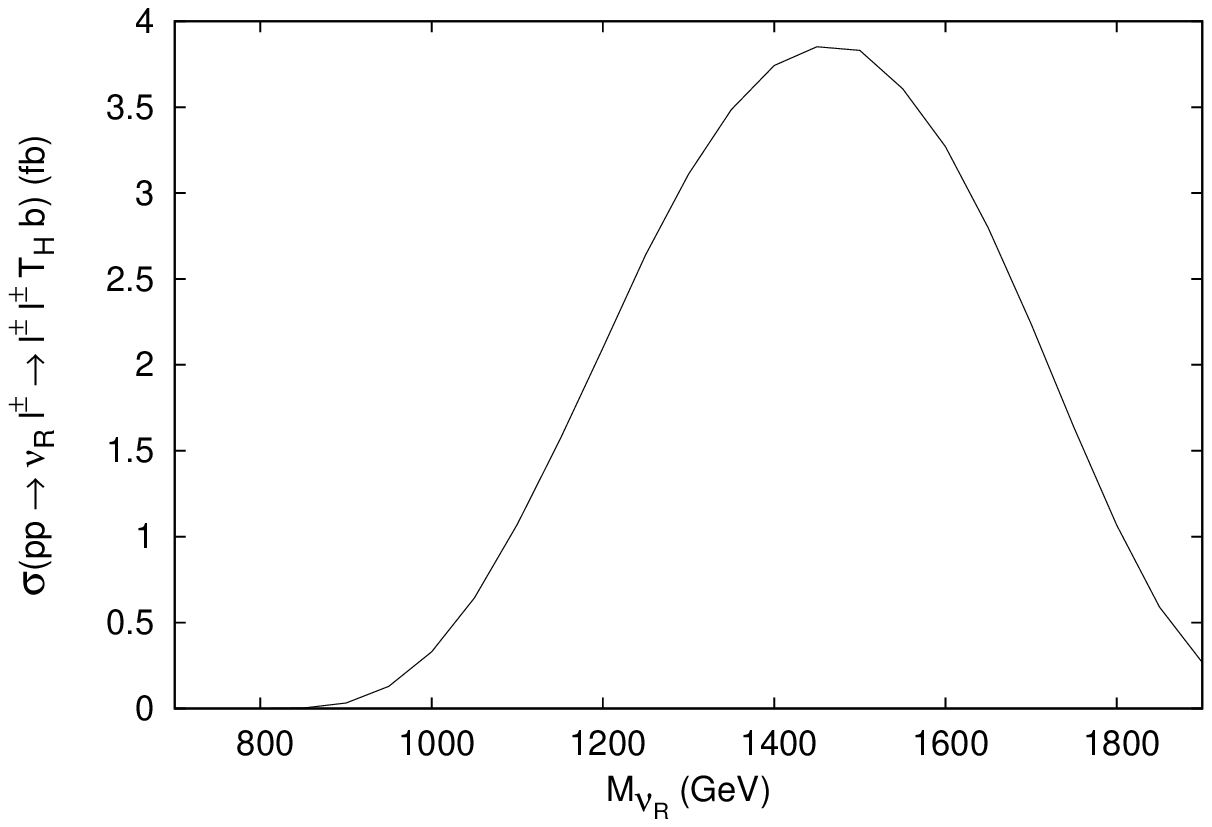}
 \caption[Production of $\nu_R l^{\pm}$ and $T_H b l l$]{The left plot shows the production cross section of the right handed neutrino $\nu_R + l^{\pm}$ with an associated lepton 
 ($e^{\pm}$ or  $\mu^{\pm}$) as a function of
 $m_{\nu_R}$. 
 The right plot shows the production cross section of the heavy top $T_H$ through the decay of $\nu_R$ 
 in association with \emph{same-sign} leptons ($e^{\pm}$ or $\mu^{\pm}$)
as a function of  $m_{\nu_R}$.} 
\label{fig:N-prod}
\end{figure}

The production cross section at the LHC of $\nu_R + l^{\pm}$, where $l=e,\mu$,
is calculated using CalcHEP \cite{calchep} and is shown in Fig.~\ref{fig:N-prod}
as a function of the right-handed neutrino mass $m_{\nu_R}$.  Half of these
events will have same-sign dilepton  without missing energy.  For example, if
$m_{\nu_R}=1$ TeV, the production cross section is about 250~fb, which leads to
approximately 3750 same-sign dilepton events with $30$ $\text{fb}^{-1}$ of
total luminosity.  The invariant mass distribution of the final state
particles should provide a clear signal of $W_R$.  Furthermore, the invariant
mass distribution of the jets plus one lepton should provide a signal of
$\nu_R$.

As mentioned above, there is no genuine SM background for the same-sign dilepton
signal and so the background at the LHC is purely instrumental.  This mostly
arises from a mismeasurement of missing energy and/or a lepton's charge.
Another possibility is the misidentification of a jet as a lepton.  
Background events can therefore arise from any of
the following processes:
%
\begin{gather}
 pp \rightarrow l^+l^-jj \label{Zdecay}   \\
 pp \rightarrow t\bar{t} \rightarrow l^+l^-b\bar{b} +  \slashed{E}_T   \label{tt} \\
pp \rightarrow W^{\pm}W^{\pm}jj \rightarrow l^{\pm}l^{\pm}jj + \slashed{E}_T \label{WWjj}\\
pp \rightarrow W^{+}W^{-}jj \rightarrow l^{+}l^{-}jj + \slashed{E}_T \label{WpWmjj}\\
pp \rightarrow t\bar{t}\rightarrow  l^{\pm}l^{\pm}bc jj + \slashed{E}_T \label{charm}\\
pp \rightarrow t\bar{t}\rightarrow l^{\pm} b\bar{b}jj + \slashed{E}_T \label{bfake}\\
pp \rightarrow jjjj \label{jjjj}\\
pp \rightarrow l^{\pm}jjj + \slashed{E}_T \label{ljjj}
\end{gather}


A complete analysis of these backgrounds requires a full detector simulation,
which is beyond the scope of this paper. Here we estimate the upper bounds of
these background processes using the parton level computation tool
Madgraph/Madevent \cite{Madgraph} to show the possible reach of the LHC.  To
approximate the detector simulation, we take the average performance of the
ATLAS detector \cite{AtlasTDR} to be 
\begin{eqnarray}
	\epsilon_l &=& 0.85 \ \ \ \  c_l  = 0.02  \\
	 R_{l,j} &=& 10^{-4}  \   R_{l,b} = 0.02,
\end{eqnarray}
where  $\epsilon_l$ is the lepton acceptance efficiency, 
$c_l$ is the probability of lepton charge misidentification, $R_{l,j}$ is the
probability that a light jet will fake a lepton and $R_{l,b}$ is the
probability that a $b$ jet will fake a lepton.  
To select events, we adopt the following basic kinematic acceptance cuts 
\begin{align}\label{basiccut}
   p_T^l>10~\textrm{GeV},  \ \ \ p_T^j>15~\textrm{GeV}, \ \ \ |\eta^{l,j}|<2.5. 
\end{align}

First consider the background events due to $l^+l^-jj$ production at the LHC as
in eq.~(\ref{Zdecay}).  The cross section for this process at the LHC is
dominated by the $Zjj$ cross section, 
which is approximately 10$^4$~pb.  After accounting for the leptonic decay of
the $Z$ and the charge mismeasurement rate, the $l^+l^-jj$ background is
approximately 40~pb, which is very large compared to the $W_R$ signal. However,
a significant reduction of this background can be easily achieved 
by applying the following additional cuts:

\begin{align}\label{morecut}
 & p_T^{leading~lepton} > 200~\textrm{GeV},\ \ \ p_T^{two \ jets} > 300~\textrm{GeV},\\ \nonumber
  & p_T^{second~lepton} > 100~\textrm{GeV}, \ \ \
    m_{ll}~>200 ~ \textrm{GeV}.
\end{align}
The cut on the invariant mass of the two leptons is useful because the two
background leptons are produced via $Z$ decay, while the signal leptons are
produced at different steps in a cascade decay.  Therefore, this cut forces the
$Z$ to be off-shell, which suppresses the background substantially.  After
imposing these additional cuts and accounting for charge misidentification, the
background cross section arising from $l^+l^-jj$ production is estimated to be
about 0.05 fb.

Background events will also arise due to $t\bar{t}$ production in which both
tops decay semi-leptonically as in eq.~(\ref{tt}).  After the basic acceptance
cuts of eq.~(\ref{basiccut}) and the extended cuts of eq.~(\ref{morecut}), we
estimate the cross section for $t\bar{t} \rightarrow b\bar{b}l^+l^- \nu_l
\bar{\nu}_l$ to be approximately 1.84~fb.  After the mismeasurement of one
lepton's charge is taken into account, this background reduces to about 0.074~fb.

The background from $W^{\pm}W^{\pm}jj$ production in eq.~(\ref{WWjj}) contains
same-sign leptons and therefore is not suppressed by the rate of charge
misdentification.  
  Note that we include both QCD and electroweak jet production in the process
  $pp \rightarrow W^{\pm}W^{\pm}jj$.
  After applying the cuts of eq.~(\ref{basiccut}) and eq.~(\ref{morecut}), the
  $l^{\pm}l^{\pm}jj$ cross section due to $W^{\pm}W^{\pm}jj$ production is
  reduced to approximately 0.07 fb.  
  
  Another background arising from $W$ production is shown in
  eq.~(\ref{WpWmjj}).  This background does not contain same-sign leptons and is
  therefore suppressed by the charge misidentification rate.  After applying the
  cuts and accounting for the probability of misidentifying the lepton's
  charge, we estimate this background cross section to be less than about
  0.02~fb.

We next consider a second background arising from $t\bar{t}$ production, as
shown in eq.~(\ref{charm}). 
Assuming two of the jets in this process (light or $b$) have $p_T > 300$~GeV,
we estimate the cross section for this process to be less than about 18~fb,
which is quite large. However, since the lepton and charm quark are typically
produced via the decay of a boosted $b$ meson, the two will be highly
collimated.  Therefore, in addition to the cuts of eq.~(\ref{basiccut}) and
eq.~(\ref{morecut}), we also require that the jet-lepton isolation, $\Delta
R_{lj}$, be greater than 0.1\footnote{Instead of studying the $b$ quark
hadronization, shower and decay, we simplify the analysis by studying the $b$
decay at the  parton level.  However, the $b$ meson produced in this reaction
is typically less energetic than the $b$ quark from the hard scattering because
energy is carried away during fragmentation. In order to account for this
difference, we consider the distribution of the scaled energy $x_E~=~E_B/E_b$
given in \cite{bmeson}.  This distribution is peaked at $x_E \sim 0.8$.  A
numerical analysis shows that taking a fixed value of $x_E~=~0.8$ is a
good approximation to the full distribution. We therefore use this scaled
energy factor in our parton level analysis.}. 
   Furthermore, we require that the invariant mass of all final state particles
   fall within 200 GeV of the mass of the $W_R$ and the total missing $p_T$
   carried by the neutrinos be less than 100~GeV.  These cuts are very
   efficient and reduce the background of eq.~(\ref{charm}) to less than
   1.84 $\times 10^{-3}$~fb.
   

In addition to producing a same-sign lepton via $b$ decay, it is also possible
for a bottom quark in eq.~(\ref{charm}) to fake a lepton, with a probability of
about 1 in 50. 
To estimate this background, we apply the lepton $p_T$ cuts of
eq.~(\ref{morecut}) to one of the $b$ jets and the jet $p_T$ cuts to two of the
remaining three jets.  After the cuts and accounting for the probability that a
$b$ jet will fake a lepton, 
we find this background can be reduced to less than about 0.55~fb.

Additional backgrounds involving the misidentification of a jet as a lepton are
shown in eqs.~(\ref{jjjj}) and (\ref{ljjj}).  We first consider the production
cross section of $jjjj$ in which two jets are misidentified as leptons.  To
determine an upper bound on the cross section for this process, we first
estimate di-jet production at the LHC, $pp\rightarrow jj$.  For $p^j_T >$ 300
GeV, the di-jet production cross section at the LHC is approximately $8.5
\times 10^3$~pb.  To pass the extended cuts of eq.~(\ref{morecut}), all of the
jets in the four-jet are required to be hard and have similar $p_T$.
Logarithmically enhanced soft radiation will not be sufficient to produce the
required high $p_T$ jets.  Therefore, four-jet background events must arise
from four partons with high $p_T$ and will be reduced by at least a factor of
$\alpha_s^2 \sim
0.015$ and two factors of the jet-lepton fake rate, $10^{-4}$.  Therefore, we
  expect the background arising from four jet events to be suppressed by at
  least a factor of $10^{-10}$ compared to the di-jet production cross section.
  Since the cuts of eq.~(\ref{morecut}) require  two jets to have $p_T
  >300$~GeV, we estimate the background cross section arising from four jets
  can be at most the di-jet production cross section with $p^j_T > 300$~GeV
  suppressed by $10^{-10}$.  Therefore, we expect the cross section for
  background events arising from the misidentification of two light jets as
  leptons to be less than approximately $1.3 \times 10^{-3}$~fb.

This type of background can also arise from $b \bar{b}jj$ production.  While
this process is suppressed compared to that of $jjjj$ production, the
probability for a $b$ jet to fake a lepton is significantly greater.  However,
since $b$ and $\bar{b}$ have opposite charge, this background is also
suppressed by the probablility of misidentifying the $b$ quark's charge.  After
accounting for this suppression, the background arising from $b\bar{b}jj$
contributes only a few percent compared to that of $jjjj$ production.  We
therefore ignore this contribution in our analysis.

%


We next consider the process $pp \rightarrow l^{\pm}\nu jjj$, in which one jet
is misidentified as a lepton as in eq.~(\ref{ljjj}).  To estimate an upper
bound on the cross section for this process, we first calculate the production
cross section of $l^{\pm}\nu jj$ requiring that both jets have $p_T > 300$~GeV
and the lepton has $p_T > 100$~GeV.  
We find this cross section to be approximately 1~pb.
After accounting for the additional hard jet with a factor of $\alpha_s$ and
the probability that a jet will fake a lepton, we find that the cross section
for $l^{\pm}l^{\pm}jj$ arising from the misidentification of a jet as a lepton
in the process $pp \rightarrow l^{\pm} \nu jjj$ to be less than approximately
0.04~fb.  As discussed previously, the light jets in this process may be
replaced by $b$ jets.  However, unlike the four-jet background, only one
$b$ jet is required to fake a lepton.  Therefore, there is no additional
suppression due to misidentifying the charge of a $b$ quark.  We estimate this
effect to contribute at most an additional 0.01~fb of background.  Therefore,
we expect the total background cross section arising from  $l^{\pm} \nu jjj$ to
be less than approximately 0.05~fb.


\begin{table}[t]
\begin{tabular}{|l|r|r|}
\hline
Process & $\sigma (p_T^j>300 \ \textrm{GeV})$ (fb) & $\sigma(p_T^j>200 \ \textrm{GeV})$ (fb) \\ \hline 
$l^+l^-jj$ & $0.05$ & $0.16$ \\ \hline
$b\bar{b}l^+l^-\nu_l\bar{\nu}_l$ & $ 0.074$& $0.25$ \\ \hline
$W^{\pm} W^{\pm}jj$ & $0.07$ & $0.13$ \\ \hline
$W^+ W^-jj$ & $0.02$ & $0.06$ \\ \hline
$l^{\pm}l^{\pm}bc\nu \nu jj$ & $ 1.8 \times 10^{-3}$ & $ 5.3 \times 10^{-3}$ \\ \hline
$l^{\pm}b\bar{b}\nu jj$ & $0.55 $ & $2.4$ \\ \hline
$jjjj$ & $1.3 \times 10^{-3}$ & $8.1 \times 10^{-3} $\\ \hline
$l^{\pm}\nu jjj$ & $ 0.05$ & $0.11$\\ \hline \hline
Total Background & $\approx 0.8 $& $ \approx 3.1$ \\ \hline \hline
Signal & 12.5 & 17 \\ \hline
\end{tabular}
\caption{A summary of the dominant backgrounds to the lepton number violating final state
$l^{\pm}l^{\pm}jj$ arising from the decay of the right-handed $W_R$.  For comparison, we 
also show the expected signal for $m_{\nu_R} = 1.5$~TeV.} 
\label{backgrounds}
\end{table}

The dominant backgrounds to the $l^{\pm}l^{\pm}jj$ signal described above are
summarized in Table~(\ref{backgrounds}).  For comparison, we also include in
this table an additional set of background cross sections in which the $p_T$
cuts on the jets are taken to be 200~GeV. We estimate the  total background in
these two cases to be less than 0.8~fb and 3.1~fb for $p_T^j > 300$~GeV and
$p_T^j >$~200~GeV, respectively.  For $m_{\nu_R} = 1.5$~TeV, the signal cross
section is about $50$~fb before the cuts and is reduced to approximately
12.5~fb (17~fb) after the cuts given that $p_T^j > 300$~GeV (200~GeV). For the
   signal, we also take into account the lepton acceptance efficiency.  
 More thorough cuts on missing energy and the invariant masses may further
 reduce these backgrounds.  However, since the expected signal of $W_R$ and
 $\nu_R$ is already much larger than the background, we do not analyze these
 backgrounds any further. 
Therefore, in this scenario it should be possible to observe the right-handed
gauge boson $W_R$ and the right-handed neutrino $\nu_R$ at the LHC.


\subsection{$T_H$ Search: $m_{\nu_R}>m_{T_H}$}

In this case, $T_H$ decays solely to $b+ \phi^{\pm}$ and results in an all jet
final state. However, $T_H$ can also be produced by the decay of $\nu_R$
through an off shell $W_R$.  The process is 
\begin{align}
  pp\rightarrow l^{\pm} \nu_R \rightarrow l^{\pm}l^{\pm}bT_H\rightarrow l^{\pm}l^{\pm}bbjj
\end{align}
with the cross section
\begin{equation}
  \sigma(pp\rightarrow T_H l^{\pm} l^{\pm} b) \approx \sigma(pp\rightarrow
  \nu_R l^{\pm}) \times Br(\nu_R \rightarrow l^{\pm}T_H b),
\end{equation}
which is a few fb, as shown in Fig.~\ref{fig:N-prod}.  By requiring that the
leptons be of the same-sign, applying kinematic cuts and possibly $b$-tagging,
it should be possible to separate these events from background at the LHC.  The
invariant mass distribution of three of the four jets should provide a signal
of $T_H$.

As with the same-sign $l^{\pm}l^{\pm}jj$ signal, the background for same-sign
$l^{\pm}l^{\pm}bbjj$ signal is purely instrumental.  To analyze the background,
we implement kinematic cuts that are identical to those of the previous section
except that we now require four hard jets,  
\begin{align}\label{jet3211}
  p_T^{two ~leading~jets} > 200~\textrm{GeV}, \ \ \ 
  p_T^{next ~two~ leading ~jets} > 80~\textrm{GeV}.
\end{align}
Since the $T_H$ signal contains at least four hard jets, background processes
similar to those described in the previous section containing only two jets
(eqs.~(\ref{Zdecay})-(\ref{WpWmjj}) and eqs.~(\ref{jjjj})-(\ref{ljjj})) should
be suppressed by  a factor of approximately $\alpha_s^2$.  This factor accounts
for the two additional hard partons required to mimic the signal.  However,
this argument does not apply to the events arising from $t\bar{t}$ production.
To satisfy the cuts, the transverse momentum of the tops must be about 400~GeV
in order to decay to a 200~GeV jet and lepton.  This means a potentially large
logarithm may be associated with the extra 80~GeV jet production due to the
large gap between 400 and 80 GeV.  Therefore, the production cross section of
$t\bar{t}$ plus jets (with $p_T > 80$~GeV) may not be suppressed compared to
that of $t\bar{t}$ production.  

The total background can be divided into two classes.  One class arises from
$t\bar{t}+X$ production, which is expected to be similar to that obtained in
the $W_R$ search and is about $2.6$ fb. The other class of background includes
everything else, which is suppressed by $\alpha_s^2$ and is about 0.007~fb. The
background is therefore dominated by $t\bar{t}+X$, which is about the size of
the signal and needs further analysis.

The same-sign lepton, four-jet background arising from $t\bar{t}$ production,
eq.~(\ref{charm}), is small compared to the signal and therefore no further
analysis is needed. However, the background arising from a $b$ quark faking a
lepton, eq.~(\ref{bfake}), and the leptonic decays of both tops, eq.~(\ref{tt})
need further suppression.  For these processes, we apply the same invariant
mass and missing $p_T$ cuts as those used in the $W_R$ search. $t\bar{t} j$ and
$t\bar{t}jj$ events are generated by MadEvent to estimate the background for
eq.~(\ref{bfake}) and eq.~(\ref{tt}). The above kinematic cuts are then imposed
on the decay products of both of the tops.  We find the cross section of
eq.~(\ref{bfake}) to be about 0.057~fb and that of eq.~(\ref{tt}) to be about
0.004~fb.

The backgrounds contributing to the $l^{\pm}l^{\pm}bbjj$ signal are summarized
in Table~(\ref{backgrounds2}).  
Given the cuts described above, we estimate the total background cross section
of the $T_H$ signal to be about  0.07~fb. To predict the possible reach of the
LHC, we must understand how the $l^{\pm}l^{\pm}bbjj$ signal is affected by the
cuts discussed above.  After the cuts, the $l^{\pm}l^{\pm}bbjj$ signal is
reduced by about 80\% when $m_{\nu_R} = 1.5$~TeV. This results in a cross
section of approximately 0.87~fb after
  accounting for the lepton acceptance efficiency.  Clearly, the signal is
  significantly larger than the estimated background. With  only 10~fb$^{-1}$
  of total luminosity, we expect about 9 signal events while the SM prediction
  is less than one.  However, due to the combinatoric ambiguity in the
  reconstruction of $T_H$, more events may be needed.  It may also be possible
  to use kinematics to pick out the correct combination.

\begin{table}[t] 
\begin{tabular}{|l|r|r|}
\hline
Background process & $\sigma(p_T^j>200 \ \textrm{GeV})$ (fb) \\ \hline 
$bcl^{\pm}l^{\pm}\nu\bar{\nu}jj$ & $0.005 $ \\ \hline
$b\bar{b}l^{\pm}\nu jjj$ & $0.057 $ \\ \hline
$b\bar{b}l^+l^-\nu\bar{\nu} jj$ & $0.004$\\ \hline
$\alpha_s^2$ Suppressed &$ 0.007  $ \\ \hline \hline
Total Background & $\approx 0.07 $ \\ \hline \hline
Signal & 0.87 \\ \hline
\end{tabular}
\caption{A summary of the dominant backgrounds to the lepton number 
violating final state $l^{\pm}l^{\pm}bbjj$ arising from the decay of a right-handed 
neutrino to a heavy top partner, $T_H$.  For comparison, we also show the 
expected signal for $m_{\nu_R} = 1.5$~TeV. 
} 
\label{backgrounds2}
\end{table}

We stress that the procedure outlined above provides only a rough estimate of
the relevant backgrounds. The true background could be larger by an $O(1)$
factor. On the other hand, a more detailed study involving missing energy cuts
and a careful utilization of invariant mass cuts 
may result in a significant improvement in the signal to background ratio.
This study, as well as the reconstruction of $T_H$, requires a more realistic
analysis, which we leave for future work. 
Therefore, we conclude that it should be possible to detect the heavy top
partner $T_H$ at the LHC, provided that $m_{\nu_R}>m_{T_H}$.



\subsection{$T_H$ Search: $m_{\nu_R} < m_{T_H}$}

To observe the leptonic decays of the heavy top in this case, we must look for
the decays of $T_H$ to $\nu_R$.  Once produced via $W_R$ decay, $T_H$ can decay
to $\nu_R +b+ l^{\pm}$ through an off shell $W_R$. 
However, there is another decay channel which does not involve $\nu_R$, $T_H
\rightarrow \phi^{\pm}+ b$.
As discussed above,  $\phi^{\pm}$ decays to jets, so the signal is either $pp
\rightarrow llbbjj$ or $pp \rightarrow bbjj$.  The cross sections for these
processes are determined by the partial decay width of $T_H$ to these channels
 \begin{eqnarray}
   \Gamma_{T_H \rightarrow \nu_R l^{\pm} b}&\sim & 10^{-5} \ \text{GeV} \nonumber \\
   \Gamma_{T_H \rightarrow b \phi^{\pm}}&\sim & 3 \ \text{GeV}. 
\end{eqnarray}
As expected, the two-body decay dominates the decay width, making the branching
fraction Br($T_H \rightarrow \nu_R b l^{\pm}$) very small.   The cross section
for $pp \rightarrow T_H b \rightarrow bblljj$ is then about $10^{-3}$ fb, which
is too small to be observed at the LHC.  Therefore, in this case it will not be
possible to detect the heavy top partner $T_H$.

\section{Conclusion}

In summary, we have shown that a TeV scale right-handed neutrino in the
left-right twin Higgs model leads to interesting lepton number violating
signatures for $W_R$ and $T_H$ at the LHC, provided that $m_{\nu_R} < m_{W_R}$.
Lepton number violating decays of right-handed $W_R$ should be observable
provided that $W_R$ and $\nu_R$ are not nearly degenerate.  Detection of the
heavy top is possible if $m_{\nu_R} > m_{T_H}$.  
These signals may be used to complement other collider searches for $W_R$ and
$T_H$.  In the limit that $M \rightarrow 0$, these signatures may be the only
way to observe the particles responsible for natural electroweak symmetry
breaking in the left-right twin Higgs.

\section{Acknowledgments} We thank Zackaria Chacko for valuable discussions and
comments on the draft. C. K is supported by the NSF under grant PHY-0801323.
H.S.G is supported by the NSF under grant PHY-04-57315 and by the DOE under
grant DE-AC02-05CH11231.

\end{document}